\documentclass[11pt,twoside]{article}
\usepackage{asp2004}
\usepackage{psfig}
\usepackage{epsf}
\usepackage{graphics}
\usepackage{lscape}
\begin{document}
\title{IR spectroscopy of pre-cataclysmic binaries}
\author{C. Tappert$^1$, B. T. G\"ansicke$^2$, R. E. Mennickent$^3$}
\affil{
$^1$Departamento de Astronom\'{\i}a y Astrof\'{\i}sica, Pontificia 
Universidad Cat\'olica, Casilla 306, Santiago 22, Chile\\
$^2$Department of Physics, University of Warwick, Coventry CV4 7AL, UK\\
$^3$Grupo de Astronom\'{\i}a, Universidad de Concepci\'on, Casilla 160-C, 
Concepci\'on, Chile
}

\begin{abstract}
We have taken near-infrared ($JHK$) spectroscopic data of a sample of pre-CVs, 
in order to examine them for signatures of nuclear evolution. Here we present 
preliminary results for the three systems BPM 71214, RR Cae, and RE 1016$-$053.
\end{abstract}
\thispagestyle{plain}

The data were obtained with SofI at the NTT, La Silla. The
low resolution Grism Blue for $J$ ($R \sim 800$) and medium resolution grisms
for $H$ ($R \sim 1200$) and $K$ ($R \sim 1300$) were used. The standard 
reduction process included correction for telluric absorption, but no flux 
calibration. Equivalent widths were measured using the wavelength ranges given 
by Wallace et al.\ (2000), Meyer et al.\ (1998), and Wallace \& Hinkle (1997).
For the $J$ spectra the continuum was determined using the regions from Wallace
et al.\ (2000), with the execption of CN. For this line and for the $H$ and $K$
data, the equivalent widths were measured with respect to linear fits to the 
local continuum. 

The results are plotted in Fig.\ 1 together with standard stars of luminosity
classes V and III. We find that within the errors there are no systematic
deviations from the values for main-sequence stars, although a few differences
for individual lines do exist. However, a proper comparison is somewhat
problematic, due to the low number of standard stars with spectral types
later than M0, and due to the low efficiency of the SofI $K$ grism longward
of 2.3$\mu$m ($<$4350 cm$^{-1}$), which practically inhibits the measurement
of those CO bands that have been found to show systematic anomulous abundances 
in CVs (Harrison et al.\ 2004). We therefore plan to supplement our survey with
$K$ spectra from ISAAC at the VLT.

\begin{figure}[hbt]
\centerline{\resizebox{11cm}{!}{\includegraphics{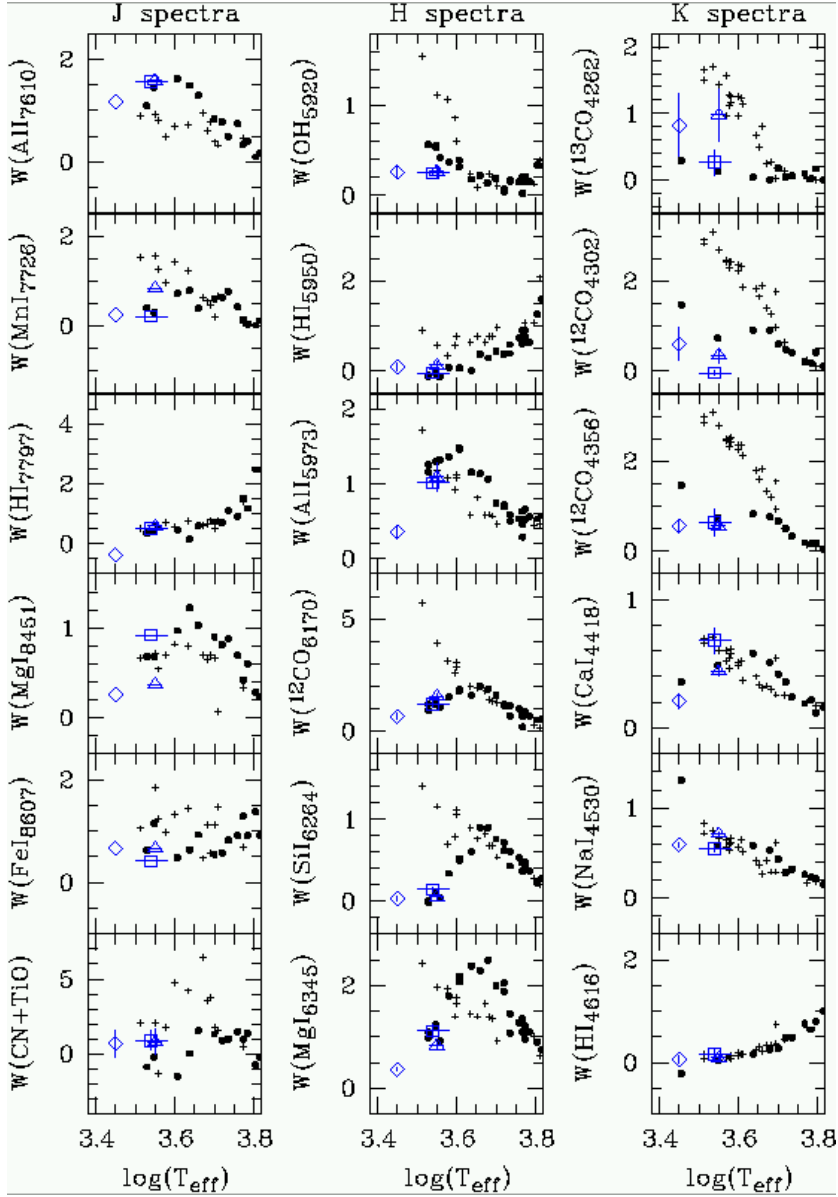}}}
\caption{Equivalent widths in cm$^{-1}$ of selected regions in the $JHK$ 
spectra. Standard star data for $J$ and $H$ have been taken from Wallace et 
al.\ (2000) and Meyer et al.\ (1998), respectively. Equivalent widths for $K$ 
standards have been measured using electronically available spectra from 
Wallace \& Hinkle (1997). In the plots, the small symbols indicate the
standard data: dots for dwarfs and crosses for giants. The pre-CVs are
symbolized as follows: Square = RE 1016-053, Triangle = BPM 71214,
Diamond = RR Cae. Horizontal error bars correspond to the uncertainty
of the spectral type given in the literature, vertical error bars give
the standard deviation of the average equivalent widths measured from
five individual fits to the local continuum.}
\end{figure}

\end{document}